\input psfig.sty

\documentstyle{new}
\begin{document}

\title{%
The Nuclear Physics of Solar and Supernova Neutrino Detection}

\author{W. C. HAXTON \\
{\it Institute for Nuclear Theory, Box 351550, and \\
Department of Physics, Box 351560, \\ 
University of Washington, Seattle, WA 98195, USA}}

\maketitle

\section*{Abstract}

This talk provides a basic introduction for students interested
in the responses of detectors to solar, supernova, and other
low-energy neutrino sources.  Some of this nuclear physics is
then applied in a discussion of nucleosynthesis within a
Type II supernova, including the r-process and the $\nu$-process.

\section{Introduction}

It is a pleasure to have this opportunity to visit Tokyo 
Metropolitan University and address this group of students and
researchers interested in neutrino physics.  Professor Minakata
has asked me to provide a pedagogical overview of the nuclear
physics governing the detection of solar, supernova, and 
other low-energy neutrinos.  As the following presentation is
very elementary, I apologize to those of you who are already
familiar with the subject.\\

The talk begins with a discussion of the allowed and first-forbidden
responses of nuclei to low-energy neutrinos.  To illustrate how
the allowed response can be crucial to efforts to detect solar
neutrinos, I discuss the classic example of the $^{37}$Cl
experiment.  Similarly, first-forbidden responses are 
generally quite important to the interaction of heavy-flavor
neutrinos from core-collapse supernovae.  I discuss some
examples from explosive nucleosynthesis -- the r-process and the
$\nu$-process -- to illustrate some of the issues. \\
  
\section{The allowed response}
Figure 1 shows several semileptonic weak interactions
that take place between nucleons or in nuclei [19].
Among such reactions important to astrophysics, two of the
most familiar are the decay of the free neutron
\[ n \rightarrow p + e^- + \bar{\nu}_e, \]
a reaction that influences the n/p ratio in big-bang nucleosynthesis,
and the driving reaction of the solar $pp$ chain
\[ p + p \rightarrow d + e^+ + \nu_e . \]
The latter can be thought of as the decay of a free proton in
the plasma, made possible energetically by the proximity
of a second proton, within the range of the
nuclear force (several fermis), so that the final n+p state
can form a bound deuteron.  It is the binding energy of the
deuteron that allows the reaction to take place. \\

\begin{figure}[htb]
\psfig{bbllx=3.5cm,bblly=4.0cm,bburx=18cm,bbury=10.0cm,figure=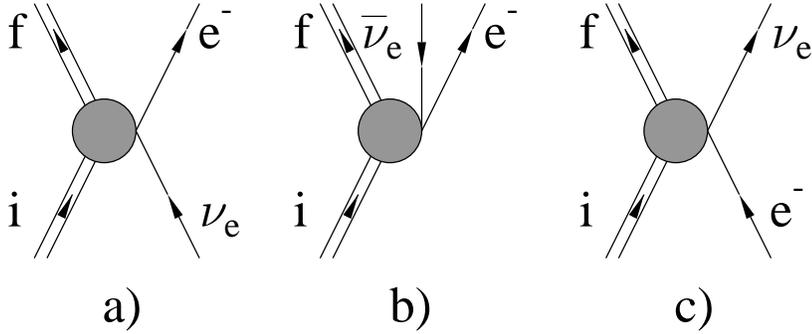,height=2.0in}
\caption{Semileptonic weak interactions of interest: a) charged
current neutrino reaction (e.g., $^{37}$Cl($\nu,e^-)^{37}$Ar);
b) $\beta^-$ decay; and c) electron capture (e.g., $^{37}$Ar(EC)$^{37}$Cl).}\
\end{figure}
  
As preparation for our discussion of nuclei, first consider the
rate for neutron $\beta$ decay
\begin{equation}
d\omega = |M|^2 {d^3p_{p} \over (2 \pi)^3} {M_p \over E_p}
{d^3p_e \over (2 \pi)^3} {m_e \over E_e} {d^3p_\nu \over (2 \pi)^3}
{m_\nu \over E_\nu} (2 \pi)^4 \delta^4(p_{n}-p_{p}-p_e-p_\nu). 
\end{equation}
(Note that I use a spinor normalization convention where all
fermions are treated as massive, including neutrinos.)
The invariant amplitude $M$ is taken 
to be a contact current-current interaction, because the
momentum transfered between the leptons and nucleon is so much
smaller than the mass of the W boson.  Thus
\begin{equation}
M = \cos \theta_c {G_F \over \sqrt{2}} \bar{u}(p)\gamma^\mu
(1-g_A\gamma_5) u(n) \bar{u}(e) \gamma_\mu (1-\gamma_5)
v(\nu) 
\end{equation}
where $G_F$ is the weak coupling constant measured in muon decay
and $\cos \theta_c$ gives the amplitude for the weak interaction
to connect the u quark to its first-generation partner, the d quark.
The origin of this effective amplitude is the underlying standard
model predictions for the elementary quark and lepton currents.
The weak interactions at this level are predicted by the standard
model to be exactly left handed.  Experiment shows that the
effective coupling of the W boson to the nucleon is governed by
\[ (1-g_A\gamma_5) \]
where $g_A \sim 1.26$.
The axial coupling is thus shifted from its underlying value
by the strong interactions responsible for the binding of the
quarks within the nucleon. \\

The extension to nuclear systems traditionally begins with the
observation that nucleons in the nucleus are rather nonrelativistic,
$v/c \sim 0.1$.  The $\beta$ decay amplitude 
$\bar{u}(p) \gamma^\mu(1-g_A \gamma_5) u(n)$ can be expanded
in powers of $p/M$.
The leading vector and axial operators
are readily found to be
\[ \begin{array}{ccc} &\underline{\gamma_\mu} & \underline{\gamma_\mu \gamma_5} \\
\mu=0 & 1 & { \vec{\sigma} \cdot \vec{p} \over M} \sim {v \over c} \\
\mu=1,2,3 & {\vec{p} \over M} \sim {v \over c} & \vec{\sigma}
\end{array} \]
Thus it is the time-like part of the vector current and the 
space-like part of the axial-vector current that survive in the
nonrelativistic limit.\\

(In a nucleus these currents must be corrected for the presence
of meson exchange contributions.  The corrections to the 
vector charge and axial three-current, which we just pointed out
survive in the nonrelativistic limit, are of order $(v/c)^2
\sim$ 1\%.  Thus the naive one-body currents
are a very good approximation to the nuclear currents.  In
contrast, exchange current corrections to the axial charge and
vector three-current operators are of order $v/c$, and thus
of relative order 1.  This difficulty for the vector three-current
can be largely circumvented, because 
current conservation as embodied in the generalized Siegert's
theorem allows one to rewrite important parts of this
operator in terms of the
vector charge operator.  In the long-wavelength limit 
appropriate to $\beta$ decay, all terms unconstrained by
current conservation do not survive.  In effect, one has
replaced a current operator with large two-body corrections
by a charge operator with only small corrections.
In contrast, the axial charge operator is significantly altered
by exchange currents even for long-wavelength processes like
$\beta$ decay.  Typical axial-charge $\beta$ decay rates are
enhanced by $\sim$ 2 because of exchange currents.) \\
  
If such a nonrelativistic reduction is done for our nucleon
$\beta$ decay amplitude, one obtains
\begin{equation}
M \sim \cos \theta_c {G_F \over \sqrt{2}} \left(
\phi^\dag(p) \phi(n) \bar{u}(e) \gamma^0(1 -\gamma_5)v(\nu) -
\phi^\dag(p) g_A \vec{\sigma} \phi(n) \cdot \bar{u}(e)
\vec{\gamma}(1-\gamma_5)v(\nu) \right) 
\end{equation}
where the $\phi$ are now two-component Pauli spinors for the nucleons.
The above result is written for the $\beta$ decay $n \rightarrow
p$.  It is convenient to generalize it for $p \leftrightarrow n$
by introducing the isospin operators
$\tau_\pm$ where $\tau_+$ $\mid$ n$\rangle$ = $\mid$ p$\rangle$ and $\tau_-$$\mid$ p$\rangle$ = $\mid$ n$\rangle$, with
all other matrix elements being zero: the free proton
does not $\beta$ decay, of course, but this is good preparation
for the generalization to nuclei.
Finally, we square the invariant amplitude, integrate over the outgoing
electron, neutrino, and final nucleon three-momenta, average
over initial nucleon spin, and sum over final nucleon spin,
electron spin, and neutrino spin.  The result is
\begin{equation}
\omega = G_F^2 \cos^2 \theta_c {1 \over 2\pi^3} 
\int_{m_e}^w (w-\epsilon)^2 \epsilon \sqrt{\epsilon^2 -m_e^2}
d\epsilon {1 \over 2} \left( |\langle f || \tau_\pm || i \rangle |^2
+g_A^2 |\langle f || \sigma \tau_\pm || i \rangle |^2 \right)
\end{equation}
where $f$ and $i$ are the final and initial nucleon states,
$w$ is the energy release in the 
decay, and $\epsilon$ is the electron energy.
The $\tau_+$ operator corresponds to $\beta^-$ decay and the
$\tau_-$ to $\beta^+$ decay.  The notation $||$ denotes a matrix element
reduced in angular momentum.  One immediately sees, for
large energy release $w$, that rates scale as $w^5$. \\

This result easily generalizes to nuclear decay.  Given our
comments about exchange currents, the first step is the
replacement
\[ \tau_\pm \rightarrow \sum_{i=1}^A \tau_\pm(i) \]
\[ \sigma \tau_\pm \rightarrow \sum_{i=1}^A \sigma(i)
\tau_\pm(i). \]
We also have to worry about an approximation in our nucleon
$\beta$ decay discussion, the treatment of the nucleon
as an elementary, structureless particle.
This is certainly appropriate for momentum scales below the
inverse size of the nucleon, as the nucleon's structure
then cannot be resolved, and for energy transfers small compared
to nucleon excitation energies.  Both conditions are easily
satisfied in neutron $\beta$ decay.
In nuclear $\beta$ decay the issue is not so clear, especially
as decays can often populate a collection of states in the 
daughter nucleus.  If the lepton states are treated as 
plane waves, the operators further generalize to
\[ \sum_{i=1}^A e^{i \vec{k} \cdot \vec{r}(i)} \tau_\pm(i) \]
\[ \sum_{i=1}^A e^{i \vec{k} \cdot \vec{r}(i)} \sigma(i)
\tau_\pm(i) \]
where $\vec{r}(i)$ is the coordinate of the $ith$ nucleon
relative to the nuclear center of mass.  (The center-of-mass
coordinate would be integrated out to give the overall
three-momentum conservation for the $\beta$ decay.)
In $\beta$ decay and in solar neutrino reactions, the three-momentum transfer to the nucleus $|\vec{k}|$
is much smaller than the typical inverse nuclear size,
$\sim 160/A^{1/3}$ MeV.  Thus as long as one is interested 
in transitions where the operators $\tau$ or $\sigma \tau$
connect the initial and final states of interest, the
effects of the momentum transfer can be ignored.  Of course,
if this is not the case, then the transition amplitude is
nonzero only because of the finite momentum transfer.  If
one expands the plane wave in powers of $\vec{k} \cdot \vec{r}(i)$,
then a transition has a degree of ``forbiddenness" according
to the number of powers required to produce a nonzero
amplitude. \\
  
For the moment we will restrict ourselves to allowed transitions
where the effects of the momentum transfer can be ignored.
The nuclear decay rate is then obtained by substituting into
the neutron result
\begin{equation}
{1 \over 2J_i+1} (|\langle f || \sum_{i=1}^A \tau_\pm (i) || i \rangle |^2
+ g_A^2 |\langle f || \sum_{i=1}^A \sigma(i) \tau_\pm(i) || i \rangle|^2).
\end{equation}
The factor $1/(2J_i+1)$ replacing the 1/2 in the neutron result 
comes from the average over initial nuclear spin directions.
As the nuclear Coulomb field can significantly distort the wave function of the outgoing
electron or positron, a final step is to correct the lepton
phase space by 
\[ F(Z,\epsilon) = |F_0(Z,\epsilon)|^2 = { 2\pi \eta \over
e^{2 \pi \eta} -1}~~\mathrm{where}~~\eta = {Z_f Z_e \alpha \over
\beta} \]
where $\beta$ is the electron/positron velocity  and 
$F_0(Z,\epsilon)$ is the s-wave Coulomb wave function in the
field of the daughter nucleus of charge $Z_f$, evaluated at
the nuclear origin.  (This is a reasonable approximation for
small $Z_f$; for heavier nuclei, however, the usual procedure
is to solve the Dirac equation for an extensive nuclear charge,
evaluating the resulting wave function at the nuclear surface.) \\

The spin-independent and spin-dependent operators appearing 
above are known as the Fermi and Gamow-Teller operators.
The Fermi operator is proportional to the isospin raising/lowering operator:
in the limit of good isopsin, which typically is good to 5\% or
better in the description of low-lying nuclear states,
it can only connect states in the same isospin multiplet,
that is, states with a common spin-spatial structure.
If the initial state has isospin $(T_i, M_{Ti})$, this final
state has $(T_i, M_{Ti} \pm 1)$ for $\beta^-$ and
$\beta^+$ decay, respectively, and is called the isospin analog state (IAS).
In the limit
of good isospin the sum rule for this operator in then
particularly simple
\begin{equation}
\sum_f {1 \over 2J_i+1} | \langle f || \sum_{i=1}^A \tau_+(i) || i \rangle |^2 =
{1 \over 2J_i+1} | \langle IAS || \sum_{i=1}^A \tau_+(i) || i \rangle |^2 = |N-Z|. 
\end{equation}
The excitation energy of the IAS relative to the parent ground
state can be estimated accurately from the Coulomb energy
difference [9]
\begin{equation}
E_{IAS} \sim ({1.728 Z \over 1.12A^{1/3} + 0.78} - 1.293) \mathrm{MeV}. 
\end{equation}
The angular distribution of the outgoing electron for a pure
Fermi $(N,Z) + \nu \rightarrow (N-1,Z+1) + e^-$ transition is 1 + $\beta \cos \theta_{\nu e}$, 
and thus forward peaked.  Here $\beta$ is the electron velocity. \\

The Gamow-Teller (GT) response is more complicated, as the 
operator can connect the ground state to many states in the
final nucleus.  In general we do not have a precise probe of 
the nuclear GT response apart from weak interactions themselves.
However a good approximate probe is provided by forward-angle
(p,n) scattering off nuclei, a technique that has been 
developed in particular by experimentalists at the Indiana
University Cyclotron Facility.  The (p,n) reaction transfers
isospin and thus is superficially like $(\nu,e^-)$.  At
forward angles (p,n) reactions
involve negligible three-momentum transfers to the nucleus.
Thus the nucleus should not be radially excited.  It thus
seems quite plausible that forward-angle (p,n) reactions
probe the isospin and spin of the nucleus, the macroscopic 
quantum numbers, and thus the Fermi and GT responses.
For typical transitions, the correspondence between (p,n) and
the weak GT operators is believed to be accurate to about 10\%.
Of course, in a specific transition, much larger discrepancies
can arise.\\

The (p,n) studies demonstrate that the GT strength tends to 
concentrate in a broad resonance 
centered at a position $\delta = E_{GT} - E_{IAS}$ relative
to the IAS given by [14]
\begin{equation}
 \delta \sim (7.0 -28.9 {N-Z \over A})~\mathrm{MeV}. 
\end{equation}
Thus while the peak of the GT resonance is substantially above the IAS for
$N \sim Z$ nuclei, it drops with increasing neutron excess.
Thus $\delta \sim 0$ for Pb.  A typical value for the full
width at half maximum $\Gamma$ is $\sim$ 5 MeV. \\

The approximate Ikeda sum rule constrains the difference
in the $\beta^-$ and $\beta^+$ strengths
\begin{equation}
\sum_f ( |M_{GT}^{fi}(\beta^-)|^2 - |M_{GT}^{fi}(\beta^+)|^2 )
= 3(N-Z)
\end{equation}
where
\begin{equation}
|M_{GT}^{fi}(\beta^-)|^2 = {1 \over 2J_i+1} 
|\langle f || \sum_{i=1}^A \sigma (i) \tau_+(i) || i \rangle |^2. 
\end{equation}
In many cases of interest in heavy nuclei, the strength in the
$\beta^+$ direction is largely blocked.  For example, in a
naive $2s1d$ shell model description of $^{37}$Cl, discussed
below, the p $\rightarrow$ n direction is blocked by the closed
neutron shell at N=20.  Thus this relation can provide an 
estimate of the total $\beta^-$ strength.  Experiment shows
that the $\beta^-$ strength found in and below the GT 
resonance does not saturate the Ikeda sum rule, typically 
accounting for $\sim (60-70)$ \% of the total.  Measured and
shell model predictions of individual GT transition strengths
tend to differ systematically by about the same factor.
Presumably the missing strength is spread over a broad interval
of energies above the GT resonance.  This is not unexpected
if one keeps in mind that the shell model is an approximate
effective theory designed to describe the long wavelength modes
of nuclei: such a model should require effective operators,
renormalized from their bare values.  Phenomenologically, the
shell model seems to require $g_A^{eff} \sim$ 1.0 as well as
a small spin-tensor term $(\sigma \otimes Y_2(\hat{r}) )_{J=1}$
of relative strength $\sim$ 0.1 [2]. \\

The angular distribution of GT $(N,Z) + \nu_e \rightarrow 
(N-1,Z+1) + e^-$ reactions is $3 - \beta \cos \theta_{\nu e}$,
corresponding to a gentle peaking in the backward direction. \\
 
The above discussion of allowed responses can be repeated for
neutral current processes such as $(\nu,\nu')$.  The analog
of the Fermi operator contributes only to elastic processes,
where the standard model nuclear weak charge is approximately
the neutron number.  As this operator does not generate
transitions, it is not yet of much interest for 
solar or supernova neutrino detection, though there are efforts
to develop low-threshold detectors (e.g., cryogenic technologies)
where the modest recoil nuclear energies might be detectable.
The analog of the GT response involves
\begin{equation}
|M_{GT}^{fi}(\nu,\nu')|^2 = {1 \over 2J_i+1}
|\langle f || \sum_{i=1}^A \sigma(i) {\tau_3(i) \over 2} || i
\rangle |^2. 
\end{equation}
The operator appearing in this expression is familiar from
magnetic moments and magnetic transitions, where the 
large isovector magnetic moment ($\mu_v \sim$ 4.706) often
leads to it dominating the orbital and isoscalar spin operators. \\

\section{The Response of the $^{37}$Cl Detector}
An interesting example of these issues in the context of a
practical detector is provided by the $^{37}$Cl solar neutrino
experiment of Davis and his collaborators.  Davis succeeded
in recovering and counting the few atoms of $^{37}$Ar 
produced by solar neutrinos in a 0.615 kiloton 
C$_2$Cl$_4$ detector via the reaction $^{37}$Cl($\nu,e^-)^{37}$Ar.
The capture rate determined from three decades of measurements
in 2.56 $\pm$ 0.16 $\pm$ 0.16 SNU [7] (1 SNU = 10$^{-36}$ captures/target atom/s),
or about 1/3 that predicted by the standard solar model, 
conventional particle physics, and the accepted value for the 
$^{37}$Cl neutrino capture cross section.  This experiment 
was the first manifestation of the solar neutrino problem and 
remains crucial to current conclusions that neutrino oscillations
may be responsible for the neutrino deficit. \\

The strong conclusions drawn from the $^{37}$Cl experiment
depend on an accurately determined neutrino capture cross
section.  Because the threshold for $^{37}$Cl$(\nu,e^-)^{37}$Ar
(0.814 MeV) is well above the pp neutrino endpoint, the
important neutrino sources are from the $^7$Be and $^8$B
solar reactions.  The $^7$Be neutrinos can only excite the
transition to the ground state of $^{37}$Ar, which is relatively
weak (log$ft$ = 5.10).  Thus the capture rate should be dominated
by the high energy $^8$B neutrinos (endpoint $\sim$ 15 MeV). \\

The nuclear (not atomic) mass difference between $^{37}$Cl 
and $^{37}$Ar is 0.303 MeV.  The Coulomb energy difference 
formula (Eq. 7) for the position of the IAS gives
\[ E_{IAS} \sim 5.22 \mathrm{MeV}. \]
So we conclude that the analog state should reside at
$\sim$ 4.92 MeV in $^{37}$Ar.  Experiment has identified the
IAS at 4.99 MeV.  In the limit of good isospin the 
superallowed (Fermi) transition to the IAS has 
$|M_F|^2$ = N - Z = 3.0; this transition accounts for about 70\%
of the $^8$B capture rate.\\

Now the interesting issue is the model-dependent GT
response.  While we have noted that the total GT response 
is about three times the Fermi response (taking $g_A^{eff}
\sim$ 1), its contribution to the capture rate depends on
its distribution, particularly at low excitation energies
where the $^8$B neutrino cross section phase space is 
large.  As the effective particle breakup threshold for $^{37}$Ar
is 8.79 MeV, GT transitions to states above this energy
clearly do not contribute.  According to our estimate (see Eq. (8))
for $\delta = E_{GT} - E_{IAS} \sim 4.66$ MeV, the peak of
the GT distribution should be at an excitation energy $\sim$ 9.6
MeV, relative to the ground state of $^{37}$Ar.
The two strongest peaks in the forward-angle (p,n) studies
are in the region between 7 and 10 MeV, roughly in accord with
expectations.  Thus much
of the GT strength is in the continuum, and still more 
resides above the IAS, where the neutrino phase space
drops rapidly with increasing excitation energy. \\

To put the capture cross section on firm ground, a reliable
map of the strength and distribution of the GT bound
state response is needed.  In 1964 Bahcall
and Barnes [4] pointed out that the needed information 
could be obtained from the delayed proton spectrum following
the $\beta$ decay of $^{37}$Ca, as illustrated in Fig. 2.
Assuming isospin invariance, the decay $^{37}$Ca($\beta^+)^{37}$K
is the mirror reaction to $^{37}$Cl($\nu,e^-)^{37}$Ar.
As the $^{37}$K levels above the first excited state are
unstable to proton emission, the allowed matrix elements for 
these levels can be deduced from the spectrum and intensities
of the delayed protons.  The transition to the ground state
of $^{37}$Ar is known directly, as this transition determines
the electron capture lifetime of $^{37}$Ar.  The final needed
constraint on the transition to the first excited state is
imposed by the total rate for $^{37}$Ca $\beta$ decay.
Thus, to the extent that isospin invariance relates the 
mirror systems accurately, the needed GT strengths 
can be taken entirely from experiment. \\

\begin{figure}[htb]
\psfig{bbllx=-3.0cm,bblly=1.5cm,bburx=18cm,bbury=22.6cm,figure=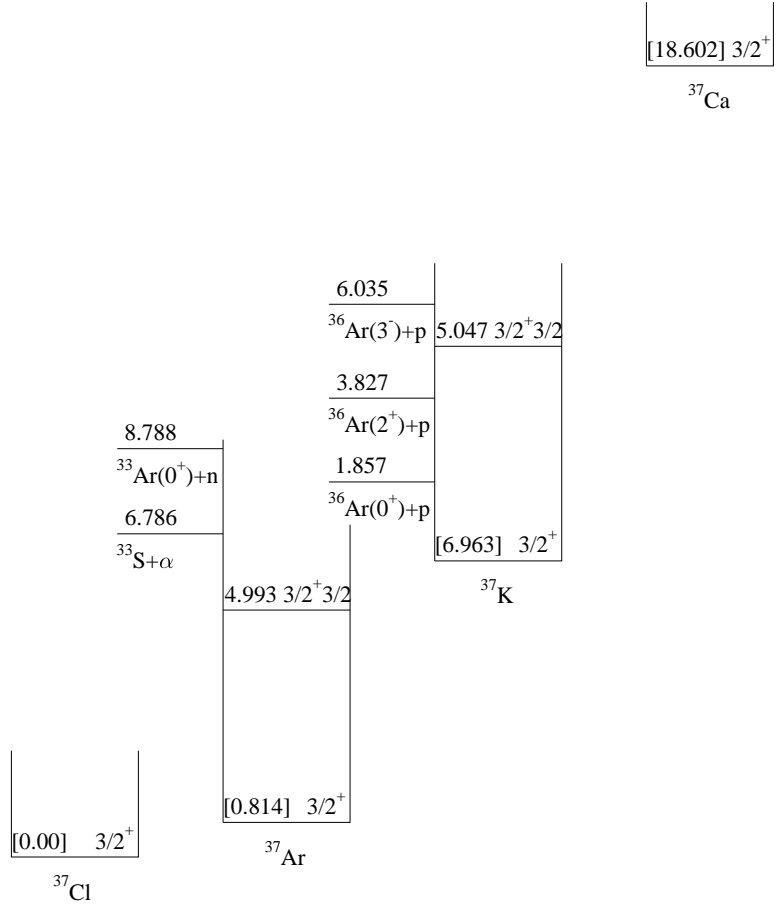,height=4.7in}
\caption{Decay schemes of A=37 nuclei.  For simplicity only the
lowest T = 1/2 and T = 3/2 levels are shown.  The important 
particle decay thresholds for states in $^{37}$Ar and $^{37}$K
are indicated.  Atomic masses in MeV are indicated by [0.00], etc.
All other energies are excitation energies in the indicated
nuclei.}
\end{figure}
  
The $^{37}$Ca($\beta^+)^{37}$K delayed proton spectrum was
measured [16]] by two groups; the deduced $ft$ values were the 
basis for the $^{37}$Cl cross section used for 20 years.
Interestingly these early experiments were flawed because
of a simplifying assumption, that the delayed protons 
from $^{37}$K were accompanied by production of the daughter
nucleus $^{36}$Ar in its ground state.  In 1987 Adelberger
and Haxton [1], noticing that the GT distribution deduced from
the delayed proton experiments differed significantly 
from that recently measured [17] in $^{37}$Cl(p,n), argued that
the likely source of this discrepancy was the population 
of $^{36}$Ar in its 2$^+$ first excited state (1.97 MeV)
in the delayed proton experiments.
The states populated in $^{37}$K by allowed $\beta$ decay
have the spins and parity 1/2$^+$, 3/2$^+$, and 5/2$^+$.
Thus the reason that the $^{36}$Ar first excited state should
be important is clear: the 3/2$^+$ and 5/2$^+$ states can
populate the 2$^+$ state by s-wave proton emission, 
while the ground state requires d-wave emission. \\

The conclusion was that the $^{37}$Ca experiment had to be
redone in a kinematically complete way, where 1.97 MeV $\gamma$s
accompanying the decay of the 2$^+$ state could be observed in
coincidence with the delayed protons.  A series of
elegant experiments were conducted by Garcia et al.[10], 
resulting in the determination $\sigma(^8$B) = 1.09 $\pm$ 0.09.
Interestingly, this value was little changed from that 
used previously: ignoring the population of the 2$^+$ 
state produced two largely compensating errors.  The affected transitions
were placed too low in energy (by 1.97 MeV), but their
strengths were also underestimated as the wrong $^{37}$Ca
$\beta$ decay phase space was then employed.  However, 
the sizeable discrepancies between the $^{37}$Ca $\beta$ decay and (p,n) mappings were 
largely resolved, thus restoring confidence that the capture
rate uncertainties in the Davis experiment were under control. \\

The reason for the discussions of this section is to illustrate
that a reliable cross section for the $^{37}$Cl experiment
was obtained only after complementary calibration techniques
were proposed, developed, and cross checked.  This careful
nuclear physics is a cornerstone of today's arguments that
the solar neutrino puzzle is likely due to new
neutrino phenomena. \\ 
  
\section{Supernovae and Supernova Neutrinos}
Consider a massive star, in excess of 10 solar masses, burning
the hydrogen in its core under the conditions of hydrostatic
equilibrium.  When the hydrogen is exhausted, the core contracts
until the density and temperature are reached where 3$\alpha \rightarrow
^{12}$C can take place.  The He is then burned to exhaustion.
This pattern (fuel exhaustion, contraction, and ignition of the 
ashes of the previous burning cycle) repeats several times,
leading finally to the explosive burning of $^{28}$Si to Fe.
For a heavy star, the evolution is rapid: the star has to work
harder to maintain itself against its own gravity, and therefore
consumes its fuel faster.  A 25 solar mass star would go through
all of these cycles in about 7 My, with the final explosive Si 
burning stage taking a few days.  The result is an 
``onion skin" structure of the precollapse star 
in which the star's history can be read by looking at the 
surface inward: there are concentric shells of H, $^4$He,
$^{12}$C, $^{16}$O and $^{20}$Ne, $^{28}$Si, and $^{56}$Fe
at the center. \\

The source of energy for this evolution is nuclear binding energy.
A plot of the nuclear binding energy $\delta$ as a function of nuclear
mass shows that the minimum is achieved at Fe.  In a scale
where the $^{12}$C mass is picked as zero:
\begin{center}
$^{12}$C~~~~~$\delta$/nucleon = 0.000 MeV \\
$^{16}$O~~~~~$\delta$/nucleon = -0.296 MeV \\
$^{28}$Si~~~~$\delta$/nucleon = -0.768 MeV \\
$^{40}$Ca~~~~$\delta$/nucleon = -0.871 MeV \\
$^{56}$Fe~~~~$\delta$/nucleon = -1.082 MeV \\
$^{72}$Ge~~~~$\delta$/nucleon = -1.008 MeV \\
$^{98}$Mo~~~~$\delta$/nucleon = -0.899 Mev
\end{center}
Thus once the Si burns to produce Fe, there is no further source
of nuclear energy adequate to support the star.  So as the last
remnants of nuclear burning take place, the core is largely
supported by degeneracy pressure, with the energy generation rate
in the core being less than the stellar luminosity.  The core
density is about 2 $\times 10^9$ g/cc and the temperature is
kT $\sim$ 0.5 MeV. \\

Thus the collapse that begins with the end of Si burning is
not halted by a new burning stage, but continues.  As gravity
does work on the matter, the collapse leads to a rapid heating
and compression of the matter.  As the nucleons in Fe are bound 
by about 8 MeV, sufficient heating can release $\alpha$s and a few
nucleons.  At the same time, the electron chemical potential is
increasing.  This makes electron capture on nuclei and any free
protons favorable,
\[ e^- + p \rightarrow \nu_e + n. \]
Note that the chemical equilibrium condition is
\[ \mu_e + \mu_p = \mu_n + \langle E_\nu \rangle. \]
Thus the fact that neutrinos are not trapped plus the rise in
the electron Fermi surface as the density increases, lead to
increased neutronization of the matter.  The escaping neutrinos carry
off energy and lepton number.  Both the electron capture and
the nuclear excitation and disassociation take energy out of the electron gas,
which is the star's only source of support.  This means that 
the collapse is very rapid.  Numerical simulations find that 
the iron core of the star ($\sim$ 1.2-1.5 solar mases) collapses
at about 0.6 of the free fall velocity [13]. \\

In the early stages of the infall the $\nu_e$s readily escape.
But neutrinos are trapped when a 
density of $\sim$ 10$^{12}$g/cm$^3$ is reached. 
At this point the neutrinos begin to scatter off the matter through
both charged current and coherent neutral current processes.  The
neutral current neutrino scattering off nuclei is particularly
important, as the scattering cross section is off the total nuclear
weak charge, which is approximately the 
neutron number.  This process transfers very little energy because
the mass energy of the nucleus is so much greater than the
typical energy of the neutrinos.  But momentum is exchanged.  
Thus the neutrino ``random walks" out of the star.  When the
neutrino mean free path becomes sufficiently short, the ``trapping
time" of the neutrino begins to exceed the time scale for the
collapse to be completed.  This occurs at a density of about
10$^{12}$ g/cm$^3$, or somewhat less than 1\% of nuclear density.
After this point, the energy released by further gravitational
collapse and the star's remaining lepton number are trapped
within the star. \\

If we take a neutron star of 1.4 solar masses and a radius of
10 km, an estimate of its binding energy is
\[ {G M^2 \over 2R} \sim 2.5 \times 10^{53} \mathrm{ergs}. \]
Thus this is roughly the trapped energy that will later be radiated in neutrinos. \\

The trapped lepton fraction $Y_L$ is a crucial parameter in the
explosion physics: a higher trapped $Y_L$ leads to a larger
homologous core, a stronger shock wave, and easier passage of
the shock wave through the outer core, as will be discussed 
below.  Most of the 
lepton number loss of an infalling mass element occurs as it
passes through a narrow range of densities just before trapping.
The reasons for this are relatively simple: on dimensional 
grounds weak rates in a plasma 
go as $T^5$, where T is the temperature.  Thus the electron capture rapidly turns on as
matter falls toward the trapping radius, and lepton number loss is
maximal just prior to trapping.  Inelastic neutrino reactions
have an important effect on these losses, as the
coherent trapping cross section goes as $E_\nu^2$ and is thus 
least effective for the lowest energy neutrinos.  As these
neutrinos escape, inelastic reactions repopulate the low
energy states, allowing the neutrino emission to continue.\\

The velocity of sound in matter rises with increasing density.
The inner homologous core, with a mass $M_{HC} \sim 0.6-0.9
$ solar masses, is that part of the iron core where the sound
velocity exceeds the infall velocity.  This allows any pressure
variations that may develop in the homologous core during infall
to even out before the collapse is completed.  As a result, the
homologous core collapses as a unit, retaining its density
profile.  That is, if nothing were to happen to prevent it, 
the homologous core would collapse to a point. \\

The collapse of the homologous core continues until nuclear
densities are reached.  As nuclear matter is rather incompressible ($\sim$ 200 MeV/f$^3$),
the nuclear equation of state is effective in halting the collapse:
maximum densities of 3-4 times nuclear are reached, e.g.,
perhaps $6 \cdot 10^{14}$ g/cm$^3$.  The innermost shell of matter
reaches this supernuclear density first, rebounds, sending a 
pressure wave out through the homologous core.  This wave
travels faster than the infalling matter, as the homologous 
core is characterized by a sound speed in excess of the infall
speed.  Subsequent shells follow.  The resulting series of pressure
waves collect near the sonic point (the edge of the homologous
core).  As this point reaches nuclear density and comes to
rest, a shock wave breaks out and begins its traversal of the 
outer core. \\

Initially the shock wave may carry an order of magnitude more energy
than is needed to eject the mantle of the star (less than 10$^{51}$
ergs).  But as the shock wave travels through the outer iron core,
it heats and melts the iron that crosses the shock front, at a 
loss of $\sim$ 8 MeV/nucleon.  The enhanced electron capture 
that occurs off the free protons left in the wake of the shock,
coupled with the sudden reduction of the neutrino opacity of
the matter (recall $\sigma_{coherent} \sim N^2$), greatly 
accelerates neutrino emission.  This is another energy loss.
[Many numerical models predict a strong ``breakout" burst of 
$\nu_e$s in the few milliseconds required for the shock wave to
travel from the edge of the homologous core to the neutrinosphere
at $\rho \sim 10^{12}$ g/cm$^3$ and $r \sim 50$ km.
The neutrinosphere is the term from the neutrino 
trapping radius, or surface of last scattering.]  The summed losses
from shock wave heating and neutrino emission are comparable to 
the initial energy carried by the shock wave.  Thus most 
numerical models fail to produce a successful ``prompt"
hydrodynamic explosion. \\

Two explosion mechanisms were seriously considered in the last
two decades.  In the prompt mechanism [8] described above, the shock wave
is sufficiently strong to survive the passage of the outer iron
core with enough energy to blow off the mantle of the star.
The most favorable results were achieved with smaller stars
(less than 15 solar masses) where there is less overlying iron,
and with soft equations of state, which produce a more compact
neutron star and thus lead to more energy release.  In part
because of the lepton number loss problems discussed earlier,
now it is widely believed that this mechanism fails for all but
unrealistically soft nuclear equations of state. \\

The delayed mechanism [5] begins with a failed hydrodynamic explosion;
after about 0.01 seconds the shock wave stalls at a radius of
200-300 km.  It exists in a sort of equilibrium, gaining energy
from matter falling across the shock front, but loosing energy
to the heating of that material.  However, after perhaps 0.5
seconds, the shock wave is revived due to neutrino heating of 
the nucleon ``soup" left in the wake of the shock.  This heating
comes primarily from charged current reactions off the nucleons
in that nucleon gas; quasielastic scattering also contributes.
This high entropy radiation-dominated gas may reach two MeV in temperature.
The pressure exerted by this gas helps to 
push the shock outward. It is important to note
that there are limits to how effective this neutrino energy 
transfer can be: if matter is too far from the core, the coupling
to neutrinos is too weak to deposite significant energy.  If too
close, the matter may be at a temperature (or soon reach a temperature)
where neutrino emission cools the matter as fast or faster than
neutrino absorption heats it.  The term
``gain radius" is used to describe the region where
useful heating is done. \\

This subject is still controversial and unclear.  The
problem is numerically challenging, forcing modelers
to handle the difficult hydrodynamics of a shock wave; the
complications of the nuclear equation of state at densities not
yet accessible to experiment; modeling in two or three dimensions;
handling the slow diffusion of neutrinos; etc.  Not all of these
aspects can be handled reasonably at the same time, even with
existing supercomputers.  Thus there is considerable disagreement
about whether we have any supernova model that succeeds in
ejecting the mantle. \\

However the explosion proceeds, there is agreement that 99\% 
of the 3 $\cdot 10^{53}$ ergs released in the collapse is 
radiated in neutrinos of all flavors.  The time scale over 
which the trapped neutrinos leak out of the protoneutron star
is about 3 seconds.  (Fits to SN1987A give, assuming an
exponential cooling $e^{-t/\tau}$, $\tau \sim$ 4.5 s [3])
Through most of their migration out of the protoneutron
star, the neutrinos are in flavor equilibrium
\[ \mathrm{e.g.},~~ \nu_e + \bar{\nu}_e \leftrightarrow \nu_\mu + \bar{\nu}_\mu. \]
As a result, there is an approximate equipartition of energy
among the neutrino flavors.  After weak decoupling, the $\nu_e$s 
and $\bar{\nu_e}$s remain in equilibrium with the matter for
a longer period than their heavy-flavor counterparts, due to 
the larger cross sections for scattering off electrons and 
because of the charge-current reactions
\[ \nu_e + n \leftrightarrow p + e^- \]
\[ \bar{\nu_e} + p \leftrightarrow n + e^+. \]
Thus the heavy flavor neutrinos decouple from deeper within the star, 
where temperatures are higher.  Typical 
calculations yield 
\[ T_{\nu_\mu} \sim T_{\nu_\tau} \sim 8 \mathrm{MeV} \]
\[ T_{\nu_e} \sim 3.5 \mathrm{MeV}~~~T_{\bar{\nu_e}} \sim 4.5 \mathrm{MeV}. \]
The difference between the $\nu_e$ and $\bar{\nu_e}$ temperatures
is a result of the neutron richness of the matter, which enhances
the rate for charge-current reactions of the $\nu_e$s, thereby keeping them coupled
to the matter somewhat longer. \\

This temperature hierarchy is crucially important to nucleosynthesis
and also to
possible neutrino oscillation scenarios.  The three-flavor MSW
level-crossing diagram is shown in Fig. 3.  One very popular
scenario attributes the solar neutrino problem to $\nu_\mu \leftrightarrow \nu_e$
transmutation; this means that a second crossing with a $\nu_\tau$
could occur at higher density.  It turns out plausible seasaw
mass patterns suggest a $\nu_\tau$ mass on the order of a few eV,
which would be interesting cosmologically.  The second crossing
would then occur outside the neutrino sphere, that is, after
the neutrinos have decoupled and have fixed spectra with the
temperatures given above.  Thus a $\nu_e \leftrightarrow \nu_\tau$ oscillation
would produce a distinctive $T \sim 8$ MeV spectrum of $\nu_e$s.
This has dramatic consequences for terrestrial detection and 
for nucleosynthesis in the supernova. \\

\begin{figure}[htb]
\psfig{bbllx=-3.0cm,bblly=4.0cm,bburx=18cm,bbury=18.0cm,figure=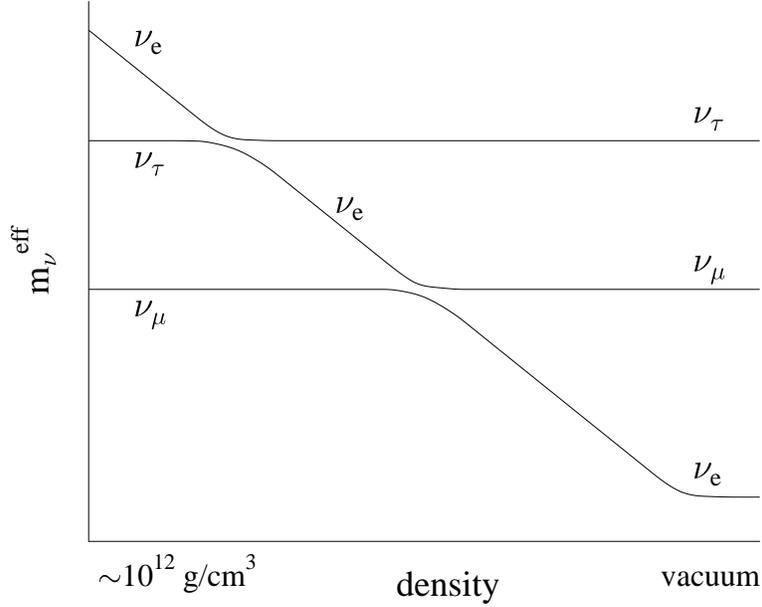,height=3.3in}
\caption{Three-flavor neutrino level-crossing diagram.  One 
popular scenario associates the solar neutrino problem with
$\nu_e \leftrightarrow \nu_\mu$ oscillations and predicts 
a cosmologically interested massive $\nu_\tau$ with 
$\nu_e \leftrightarrow \nu_\tau$ oscillations near the
supernova neutrinosphere.}
\end{figure}
  
\section{First Forbidden Responses and the Neutrino Process}
Core-collapse supernovae are one of the
major engines driving galactic chemical evolution, producing 
and ejecting the metals that enrich our galaxy.  The discussion
of the previous section described the hydrostatic evolution of
a presupernova star in which large quantities of the most
abundant metals (C, O, Ne, ...) are synthesized and later
ejected during the explosion.  During the passage of the
shock wave through the star's mantle, temperature of $\sim (1-3) \cdot 10^9$K and
are reached in the silicon, oxygen, and neon shells.  This
shock wave heating induces $(\gamma,\alpha) \leftrightarrow 
(\alpha,\gamma)$ and related reactions that generate a 
mass flow toward highly bound nuclei, resulting in the
synthesis of iron peak elements as well as less abundant
odd-A species.  Rapid neutron-induced reactions are thought
to take place in the high-entropy atmosphere just above
the mass cut, producing about half of the heavy elements 
above A $\sim$ 80.  This is the subject of the next section.
Finally, the $\nu$-process described below is responsible 
for the synthesis of rare species such as $^{11}$B and $^{19}$F.
This process involves the response of nuclei at momentum transfers
where the allowed approximation is no longer valid.  Thus we
will use the $\nu$-process in this section to illustrate some of
the relevant nuclear physics. \\

One of the problems -- still controversial -- that may be connected
with the neutrino process is
the origin of the light elements Be, B and Li, elements which are
not produced in sufficient amounts in the big bang or in any of
the stellar mechanisms we have discussed.
The traditional explanation has been cosmic ray spallation interactions
with C, O, and N in the interstellar medium.  In this picture,
cosmic ray protons collide with C at relatively high energy,
knocking the nucleus apart.  So in the debris one can find 
nuclei like $^{10}$B, $^{11}$B, and $^7$Li.\\

But there are some problems with this picture.  First of all,
this is an example of a secondary mechanism: the interstellar
medium must be enriched in the C, O, and N to provide the 
targets for these reactions.  Thus cosmic ray spallation must 
become more effective as the galaxy ages.  The 
abundance of boron, for example, would tend to grow 
quadratically with metalicity, since the rate of production
goes linearly with metalicity.  But
observations, especially recent measurements with the Hubble
Space Telescope (HST), find a linear growth in the boron abundance [18]. \\

A second problem is that the spectrum of cosmic ray protons
peaks near 1 GeV, leading to roughly comparable production of the
two isotopes $^{10}$B and $^{11}$B.  That is, while it takes 
more energy to knock two nucleons out of carbon than one, this
difference is not significant compared to typical cosmic ray
energies.  More careful studies
lead to the expectation that the abundance ratio
of $^{11}$B to $^{10}$B might be $\sim$ 2.  In nature, it is
greater than 4. \\

Fans of cosmic ray spallation have offered solutions to these
problems, e.g., similar reactions occurring in the atmospheres
of nebulae involving lower energy cosmic rays.  
As this suggestion was originally stimulated by the observation of nuclear
$\gamma$ rays from Orion, now retracted, some of the motivation
for this scenario has evaporated.  Here I 
focus on an alternative explanation, synthesis via neutrino spallation.\\

Previously we spoke about weak interactions in nuclei involving
the Gamow-Teller (spin-flip) and Fermi operators.  These are
the appropriate operators when one probes the nucleus at
a wavelength -- that is, at a size scale -- where the nucleus
responds like an elementary particle.  We can then 
characterize its response by its macroscopic quantum numbers,
the spin and charge.  On the other hand, the nucleus is a
composite object and, therefore, if it is probed at shorter
length scales, all kinds of interesting radial excitations will
result, analogous to the vibrations of a drumhead.  
For a reaction like neutrino scattering off a nucleus, the
full operator involves the additional factor
\[ e^{i \vec{k} \cdot \vec{r}} \sim 1 + i \vec{k} \cdot \vec{r} \]
where the expression on the right is valid if the magnitude of
$\vec{k}$ is not too large.  Thus the full charge operator 
includes a ``first forbidden" term
\[ \sum_{i=1}^A \vec{r}_i \tau_3(i) \]
and similarly for the spin operator
\[ \sum_{i=1}^A [\vec{r}_i \otimes \vec{\sigma}(i)]_{J=0,1,2} \tau_3(i). \]
These operators generate collective radial excitations,
leading to the so-called ``giant resonance" excitations in nuclei.
The giant resonances are typically at an excitation energy of
20-25 MeV in light nuclei.  One important property is that these
operators satisfy a sum rule (Thomas-Reiche-Kuhn) of the form
\[ \sum_f | \langle f | \sum_{i=1}^A r(i) \tau_3(i) | i \rangle |^2
\sim {N Z \over A} \sim {A \over 4} \]
where the sum extends over a complete set of final nuclear states.
These first-forbidden operators tend to dominate the cross sections
for scattering the high energy supernova neutrinos ($\nu_{\mu}$s 
and $\nu_\tau$s), with $E_\nu \sim$ 25 MeV, off light nuclei.
From the sum rule above, it follows that nuclear cross sections per
target {\it nucleon} are roughly constant. \\

The E1 giant dipole mode described above is depicted qualitatively
in Fig. 4a.  This description, which corresponds to an early model
of the giant resonance response by Goldhaber and Teller [11], 
involves the harmonic oscillation of the proton and neutron 
fluids against one another.  The restoring force for small 
displacements would be linear in the displacement and 
dependent on the nuclear symmetry energy.  There is a natural
extension of this model to weak interactions, where axial
excitations occur.  For example, one can envision a mode
similar to that of Fig. 4a where
the spin-up neutrons and spin-down protons oscillate against
spin-down neutrons and spin-up protons, the spin-isospin mode
of Fig. 4b.  This mode is one that arises in a
simple SU(4) extension of the Goldhaber-Teller model,
derived by assuming that the nuclear force is spin and isospin
independent, at the same excitation energy as the E1 mode.
In full, the Goldhaber-Teller model predicts a degenerate 15-dimensional supermultiplet of 
giant resonances, each obeying sum rules analogous to 
the TRK sum rule.  While more sophisticated descriptions of the
giant resonance region are available, of course, this crude
picture is qualitatively accurate. \\
  
\begin{figure}[htb]
\psfig{bbllx=0.0cm,bblly=3.5cm,bburx=18cm,bbury=10.0cm,figure=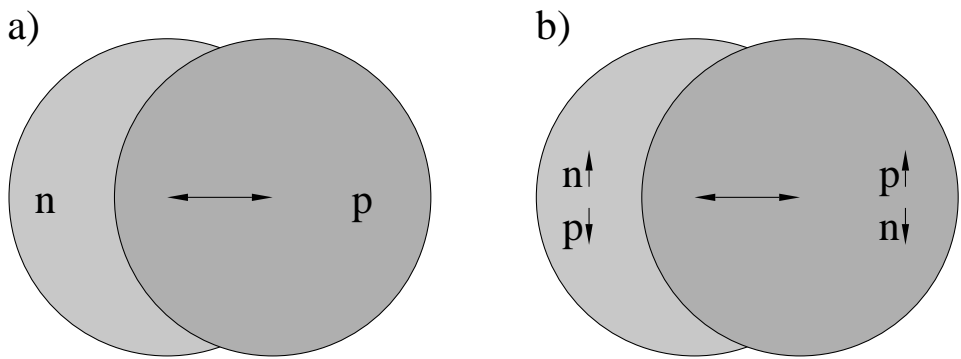,height=1.8in}
\caption{Schematic illustration of a) the E1 giant dipole mode
familiar from electromagnetic interactions and b) a spin-isospin
giant dipole mode associated with the first-forbidden weak
axial response.}
\end{figure}
  
This nuclear physics is important to the $\nu$-process [21].
The simplest example of $\nu$-process nucleosynthesis involves the Ne shell
in a supernova.  Because of the first-forbidden contributions,
the cross section for inelastic neutrino scattering to the 
giant resonances in Ne is $\sim 3 \cdot 10^{-41}$ cm$^2$/flavor
for the more energetic heavy-flavor neutrinos.
This reaction
\[ \nu + A \rightarrow \nu' + A^* \]
transfers an energy typical of giant resonances, $\sim$ 20 MeV.
A supernova releases about 3 $\times 10^{53}$ ergs
in neutrinos, which converts to about $4 \times 10^{57}$ heavy
flavor neutrinos.  The Ne shell in a 20 M$_\odot$ star has
at a radius $\sim$ 20,000 km.  Thus the neutrino fluence through
the Ne shell is
\[ \phi \sim { 4 \cdot 10^{57} \over 4 \pi (20,000 \mathrm{km})^2 }
\sim 10^{38}/\mathrm{cm}^2. \]
Thus folding the fluence and cross section,
one concludes that approximately 1/300th of the Ne nuclei interact.\\

This is quite interesting since the astrophysical origin of $^{19}$F
had not been understood.  The only stable isotope of fluorine,
$^{19}$F has an abundance
\[ {^{19}\mathrm{F} \over ^{20}\mathrm{Ne}} \sim {1 \over 3100}. \]
This leads to the conclusion that the fluorine 
found in toothpaste was
created by neutral current neutrino reactions deep inside some
ancient supernova. \\

The calculation [21] of the final $^{19}$F/$^{20}$Ne ratio is 
more complicated than the simple 1/300 ratio given above: \\
$\bullet$ When Ne is excited by $\sim$ 20 MeV through inelastic
neutrino scattering, it breaks up in two ways
\[ ^{20}\mathrm{Ne}(\nu,\nu')^{20}\mathrm{Ne}^* \rightarrow ^{19}\mathrm{Ne} + n 
\rightarrow ^{19}\mathrm{F} + e^+ + \nu_e + n \]
\[ ^{20}\mathrm{Ne}(\nu,\nu')^{20}\mathrm{Ne}^* \rightarrow ^{19}\mathrm{F}
+ p \]
with the first reaction occurring half as frequently as the 
second.  As both channels lead to $^{19}$F, we have correctly
estimated the instantaneous abundance ratio in the Ne shell of
\[ {^{19}\mathrm{F} \over ^{20}\mathrm{Ne}} \sim {1 \over 300}. \]
$\bullet$ We must also address the issue of whether the produced $^{19}$F
survives.  In the first 10$^{-8}$ sec the coproduced neutrons in
the first reaction react via
\[ ^{15}\mathrm{O}(n,p)^{15}\mathrm{N}~~^{19}\mathrm{Ne}(n,\alpha)^{16}\mathrm{O}~~
^{20}\mathrm{Ne}(n,\gamma)^{21}\mathrm{Ne}~~^{19}\mathrm{Ne}(n,p)^{19}\mathrm{F} \]
with the result that about 70\% of the $^{19}$F produced via
spallation of neutrons is then immediate destroyed, primarily
by the $(n,\alpha)$ reaction above.  In the next $10^{-6}$ sec
the coproduced protons are also processed
\[ ^{15}\mathrm{N}(p,\alpha)^{12}\mathrm{C}~~^{19}\mathrm{F}(p,\alpha)^{16}\mathrm{O}~~
^{23}\mathrm{Na}(p,\alpha)^{20}\mathrm{Ne} \]
with the latter two reactions competing as the primary proton
poisons.  This makes an important prediction: stars with high Na
abundances should make more F, as the $^{23}$Na acts as a proton
poison to preserve the produced F.\\
$\bullet$ Finally, there is one other destruction mechanism, the
heating associated with the passage of the shock wave.  It
turns out the the F produced prior to shock wave passage can
survive if it is in the outside half of the Ne shell.  The reaction
\[ ^{19}\mathrm{F}(\gamma,\alpha)^{15}\mathrm{N} \]
destroys F for peak explosion temperatures exceeding $1.7 \cdot 10^9$K.
Such a temperature is produced at the inner edge of the Ne 
shell by the shock wave heating, but not at the outer edge.\\

If all of this physics in handled is a careful network code that
includes the shock wave heating and F production both before and
after shock wave passage, the following are the results:
\[ \begin{array}{cc} \underline{[^{19}\mathrm{F}/^{20}\mathrm{Ne}]/
[^{19}\mathrm{F}/^{20}\mathrm{Ne}]_\odot} & \underline{T_{\mathrm{heavy}~\nu} \mathrm{(MeV)}} \\
0.14 & 4 \\ 0.6 & 6 \\ 1.2 & 8 \\ 1.1 & 10 \\ 1.1 & 12 \end{array} \]
One sees that the attribution of F to the neutrino process argues
that the heavy flavor $\nu$ temperature must be greater than 6 MeV,
a result theory favors.  One also sees that F cannot be overproduced
by this mechanism: although the instantaneous production of F
continues to grow rapidly with the neutrino temperature, too
much F results in its destruction through the $(p,\alpha)$
reaction, given a solar abundance of the competing proton poison
$^{23}$Na.  Indeed, this illustrates an odd quirk: although 
in most cases the neutrino process is a primary mechanism, one needs
$^{23}$Na present to produce significant F. Thus in this case the neutrino
process is a secondary mechanism. \\

While there are other significant neutrino process products ($^7$Li,
$^{138}$La, $^{180}$Ta, $^{15}$N ...), the most important 
product is $^{11}$B, produced by spallation off carbon.
A calculation by Timmes et al. [18] found that the combination of
the neutrino process, cosmic ray spallation and big-bang 
nucleosythesis together can explain the evolution of the light
elements.  The neutrino process, which produces a great deal 
of $^{11}$B but relatively little $^{10}$B, combines with the
cosmic ray spallation mechanism to yield the observed
isotope ratio.  Again, one prediction of this picture is that
early stars should be $^{11}$B rich, as the neutrino process 
is primary and operates early in our galaxy's history; the
cosmic ray production of $^{10}$B is more recent.
There is hope that HST studies will soon be able to descriminate
between $^{10}$B and $^{11}$B: as yet this has not been done. \\

\section{The r-process}
Beyond the iron peak nuclear Coulomb barriers become so high
that charged particle reactions become ineffective, leaving
neutron capture as the mechanism responsible for producing
the heaviest nuclei.
If the neutron abundance is modest,
this capture occurs in such a way that each newly synthesized
nucleus has the opportunity to $\beta$ decay, if it is energetically
favorable to do so.  Thus weak equilibrium is maintained within
the nucleus, so that synthesis is along the path of stable 
nuclei.  This is called the s- or slow-process.  However a
plot of the s-process in the (N,Z) plane reveals that this
path misses many stable, neutron-rich nuclei that are known to
exist in nature.  This suggests that another mechanism is at
work, too.  Furthermore, the abundance peaks found in nature 
near masses A $\sim$ 130 and A $\sim$ 190, which mark the closed
neutron shells where neutron capture rates and $\beta$ decay
rates are slower, each split into two subpeaks.  One set of subpeaks
corresponds to the closed-neutron-shell numbers N $\sim$ 82
and N $\sim$ 126, and is clearly associated with the s-process.
The other set is shifted to smaller N, $\sim$ 76 and $\sim$ 116,
respectively, and is suggestive of a much more explosive
neutron capture environment where neutron capture can be
rapid. \\
  
This second process is the r- or rapid-process, characterized by: \\
$\bullet$ The neutron capture is fast compared to $\beta$ decay rates. \\
$\bullet$ The equilibrium maintained within a nucleus is established by $(n,\gamma) \leftrightarrow
(\gamma,n)$: neutron capture fills up the available bound levels in
the nucleus until this equilibrium sets in.  The new Fermi level
depends on the temperature and the relative $n/\gamma$ abundance.\\
$\bullet$ The nucleosynthesis rate is thus controlled by the $\beta$
decay rate: each $\beta^-$ capture coverting n $\rightarrow$ p 
opens up a hole in the neutron Fermi sea, allowing another neutron
to be captured. \\
$\bullet$ The nucleosynthesis path is along exotic, neutron-rich
nuclei that would be highly unstable under normal laboratory conditions. \\
$\bullet$ As the nucleosynthesis rate is controlled by the $\beta$
decay, mass will build up at nuclei where the $\beta$ decay rates
are slow.  It follows, if the neutron flux is reasonable steady 
over time so that equilibrated mass flow is reached, that the
resulting abundances should be inversely proportional to these
$\beta$ decay rates. \\
  
Let's first explore the $(n,\gamma) \leftrightarrow (\gamma,n)$
equilibrium condition, which requires that the rate for $(n,\gamma)$
balances that for $(\gamma,n)$ for an average nucleus.
So consider the formation cross section
\[ A + n \rightarrow (A+1) + \gamma \]
This is an exothermic reaction, as the neutron drops into the
nuclear well.  Our averaged cross section, assuming a resonant
reaction (the level density is high in heavy nuclei) is
(see any standard nuclear astrophysics text, such as Clayton [6])
\begin{equation}
\langle \sigma v \rangle_{(n,\gamma)} = 
\left( {2 \pi \over \mu kT} \right)^{3/2} {\Gamma_n \Gamma_\gamma
\over \Gamma} e^{-E/KT} 
\end{equation}
where E $\sim$ 0 is the resonance energy,
and the $\Gamma$s are the indicated partial and total widths.
Thus the rate per unit volume is
\begin{equation}
r_{(n,\gamma)} \sim N_n N_A \left( {2 \pi \over \mu kT} \right)^{3/2}
{\Gamma_n \Gamma_\gamma \over \Gamma}
\end{equation}
where $N_n$ and $N_A$ are the neutron and nuclear number densities
and $\mu$ the reduced mass.
This has to be compared to the $(\gamma,n)$ rate. \\

The $(\gamma,n)$ reaction requires the photon number density in
the gas.  This is given by the Bose-Einstein distribution
\begin{equation}
N(\epsilon) = {8 \pi \over c^3 h^3} {\epsilon^2 d \epsilon
\over e^{\epsilon/kT} -1 }
\end{equation}
The high-energy tail of the normalized distribution can thus
be written
\[ \sim {1 \over N_\gamma \pi^2} \epsilon^2 e^{-\epsilon/kT} d \epsilon \]
where in the last expression we have set $\hbar = c = 1$. \\

Now we need the resonant cross section in the $(\gamma,n)$ 
direction.  For photons the wave number is proportional to
the energy, so
\begin{equation}
\sigma_{(\gamma,n)} = {\pi \over \epsilon^2}
{\Gamma_\gamma \Gamma_n \over (\epsilon-E_r)^2 + (\Gamma/2)^2 }
\end{equation}
As the velocity is c =1,
\begin{equation}
\langle \sigma v \rangle = {1 \over \pi^2 N_\gamma}
\int_0^\infty \epsilon^2 e^{-\epsilon/kT} d \epsilon
{\pi \over \epsilon^2} {\Gamma_\gamma \Gamma_n \over 
(\epsilon-E_r)^2 +(\Gamma/2)^2} 
\end{equation}
We evaluate this in the usual way for a sharp resonance,
remember that the energy integral over just the denominator
above (the sharply varying part) is $2 \pi/ \Gamma$:
\[ \sim {\Gamma_\gamma \Gamma_n \over N_\gamma} e^{-E_r/kT}
{2 \over \Gamma} \]
So that the rate becomes
\begin{equation}
r_{(\gamma,n)} \sim 2 N_{A+1} {\Gamma_\gamma \Gamma_n
\over \Gamma} e^{-E_r/kT} 
\end{equation}
Equating the $(n,\gamma)$ and $(\gamma,n)$ rates and taking
$N_A \sim N_{A-1}$ then yields
\begin{equation}
N_n \sim {2 \over (\hbar c)^3} \left( {\mu c^2 kT \over
2 \pi} \right)^{3/2} e^{-E_r/kT} 
\end{equation}
where the $\hbar$s and $c$s have been properly inserted to give
the right dimensions.  Now $E_r$ is esssentially the binding
energy.  So plugging in the conditions $N_n \sim 3 \times 10^{23}$/cm$^3$
and $T_9 \sim 1$, we find that the binding energy is 
$\sim$ 2.4 MeV.  Thus neutrons are bound by about 30 times $kT$,
a value that is still small compared to a typical
binding of 8 MeV for a normal nucleus.  (In this calculation
I calculated the neutron reduced mass assuming a nuclear target
with A=150.) \\

The above calculation fails to count spin states for the photons
and nuclei and is thus not quite correct.  But it makes the
essential point: the r-process involves very exotic species
largely unstudied in any terrestrial laboratory.  It is good
to bear this in mind, as in the following section we will 
discuss the responses of such nuclei to neutrinos.  Such responses
thus depend on the ability of theory to extrapolate responses
from known nuclei to those quite unfamiliar. \\

The path of the r-process is along neutron-rich nuclei, 
where the neutron Fermi sea is just $\sim$ (2-3) MeV away from
the neutron drip line (where no more bound neutron levels exist).
After the r-process finishes (the neutron exposure ends)
the nuclei decay back to the valley of stability by $\beta$
decay.  This can involve some neutron spallation ($\beta$-delayed
neutrons) that shift the mass number A to a lower value.
But it certainly involves conversion of neutrons into protons,
and that shifts the r-process peaks at N $\sim$ 82 and 126
to a lower N, off course.  This effect is clearly seen in the
abundance distribution: the r-process peaks are shifted to
lower N relative to the s-process peaks.  This is the origin of the 
second set of ``subpeaks" mentioned at the start of the section. \\

It is believed that the r-process can proceed to very heavy
nuclei (A $\sim$ 270) where it is finally ended by $\beta$-delayed
and n-induced fission, which feeds matter back into the
process at an A $\sim$ A$_{max}$/2.  Thus there may be important
cycling effects in the upper half of the r-process distribution.\\
  
What is the site(s) of the r-process?  This has been debated 
many years and still remains a controversial subject.\\
$\bullet$ The r-process requires exceptionally explosive conditions 
\begin{center}
$\rho$(n) $\sim 10^{20}$ cm$^{-3}$~~~T $\sim 10^9$K~~~t $\sim$ 1s.
\end{center}
$\bullet$ Both primary and secondary sites proposed. 
Primary sites are those not requiring preexisting metals.
Secondary sites are those where the neutron capture occurs
on preexisting s-process seeds.\\
$\bullet$ Suggested primary sites include the
the neutronized atmosphere above the proto-neutron star in
a Type II supernova, neutron-rich jets produced in supernova
explosions or in neutron star mergers, inhomogeneous big
bangs, etc. \\
$\bullet$ Secondary sites, where $\rho$(n) can be lower for 
successful synthesis, include the He and C zones in Type II
supernovae, the red giant He flash, etc.\\

The balance of evidence favors a primary site, so one requiring
no preenrichment of heavy s-process metals.  Among the evidence: \\
  
\noindent
1) HST studies of very-metal-poor halo stars: 
The most important evidence are the recent HST measurements of 
Sneden et al. [15] of very metal-poor stars ([Fe/H] $\sim$ -1.7 to -3.12)
where an r-process distribution very much like that of our sun
has been seen for Z $\gsim$ 56.  Furthermore, in these stars
the iron content is variable.  This suggests that the ``time
resolution" inherent in these old stars is short compared to
galactic mixing times (otherwise Fe would be more constant).
The conclusion is that the r-process material in these stars
is most likely from one or a few local supernovae.  The fact
that the distributions match the solar r-process (at least 
above charge 56) strongly suggests that there is some kind of
unique site for the r-process: the solar r-process distribution
did not come from averaging over many different kinds of
r-process events.  Clearly the fact that these old stars are
enriched in r-process metals also strongly argues for a 
primary process: the r-process works quite well in an
environment where there are few initial s-process metals.\\

\noindent
2) There are also fairly good theoretical arguments that a primary
r-process occurring in a core-collapse supernova might be
viable [20].  First, galactic chemical evolution studies indicate that 
the growth of r-process elements in the galaxy is consistent 
with low-mass Type II supernovae in rate and distribution.
More convincing is the fact that modelers have shown that the
conditions needed for an r-process (very high neutron densities,
temperatures of 1-3 billion degrees) might be realized in a
supernova.  The site is the last material blown off the 
supernova, the material just above the mass cut.  When
this material is blown off the star initially, it is a very
hot neutron-rich, radiation-dominated gas containing neutrons
and protons, but an excess of the neutrons.  As it expands
off the star and cools, the material first goes through
a freezeout to $\alpha$ particles, a step that essentially
locks up all the protons in this way.
Then the $\alpha$s interact through reactions like 
\[ \alpha + \alpha +\alpha \rightarrow ^{12}C \]
\[ \alpha + \alpha + n \rightarrow ^9Be \]
to start forming heavier nuclei.  Note, unlike the big bang,
that the density is high enough to allow such three-body 
interactions to bridge the mass gaps at A = 5,8.  The
$\alpha$ capture continues up to heavy nuclei,
to A $\sim$ 80, in the network calculations.  
The result is a small number of ``seed" nuclei,
a large number of $\alpha$s, and excess neutrons.  These 
neutrons preferentially capture on the heavy seeds to
produce an r-process.  Of course, what is necessary is to
have $\sim$ 100 excess neutrons per seed in order to 
successfully synthesize heavy mass nuclei.  Some of the
modelers find conditions where this almost happens. \\
  
There are some very nice aspects of this site: the amount of
matter ejected is about 10$^{-5} - 10^{-6}$ solar masses,
which is just about what is needed over the lifetime of the
galaxy to give the integrated r-process metals we see,
taking a reasonable supernova rate.  But there are also
a few problems, especially the fact that with calculated entropies
in the nucleon soup above the proto-neutron star, neutron fractions
appear to be too low to produce a successful A $\sim$ 190 peak.
There is some interesting recent work invoking neutrino oscillations
to cure this problem: charge current reactions on free protons
and neutrons determine the n/p ratio in the gas.  Then, for example, an oscillation
of the type $\bar{\nu}_e \rightarrow \nu_{\mathrm{sterile}}$ can alter this
ratio, as it would turn off the $\nu_e$s that destroy neutrons
by charged-current reactions.  Unfortunately, 
a full discussion of such possibilities would take
us too far afield today. \\

The nuclear physics of the r-process tells us that the synthesis
occurs when the nucleon soup is in the temperature range of
(3-1) $\cdot 10^9$K, which, in the hot bubble r-process described above, corresponds to a freezeout radius of
(600-100) km and a time $\sim$ 10 seconds after core collapse.
The neutrino fluence after freezeout (when the temperature
has dropped below 10$^9$K and the r-process stops) is then $\sim$
(0.045-0.015) $\cdot 10^{51}$ ergs/(100km). 
Thus, after completion of the r-process, the newly synthesized
material experiences an intense flux of neutrinos.
This brings up the question of whether the neutrino flux could
have any effect on the r-process.  

\section{Neutrinos and the r-process}
Rather than describe the exotic effects of neutrino oscillations
on the r-process, mentioned briefly above, we will examine
standard-model effects that are nevertheless quite interesting.
The nuclear physics of this section -- neutrino-induced neutron
spallation reactions -- is also relevant to recently proposed
supernova neutrino observatories such as OMNIS and LAND.
In contrast to our first discussion of the $\nu$-process in
Section 5, it is apparent that neutrino effects could be much
larger in the hot bubble r-process: the synthesis
occurs {\it much} closer to the star than our Ne radius of
20,000 km: estimates are 600-1000 km.  The r-process is completed
in about 10 seconds (when the temperature drops to about 
one billion degrees), but the neutrino flux is still significant
as the r-process freezes out.  The net result is that the
``post-processing" neutrino fluence - the fluence that can
alter the nuclear distribution after the r-process is completed -
is about 100 times larger than that responsible for fluorine
production in the Ne zone.  Recalling that 1/300 of the nuclei
in the Ne zone interacted with neutrinos, and remembering that
the relevant neutrino-nucleus cross sections scale as A, one
quickly sees that the probability of a r-process nucleus 
interacting with the neutrino flux is approximately unity.\\

Because the hydrodynamic conditions of the r-process are highly
uncertain, one way to attack this problem is to work backward
in time.  We know the final r-process distribution (what nature
gives us) and we can calculate neutrino-nucleus interactions
relatively well.  Thus from the observed r-process distribution
(including neutrino postprocessing) we can work backward to
find out what the r-process distribution looked like at the
point of freezeout.  In Figs. 5 and 6, the ``real" r-process
distribution - that produced at freezeout - is given by the 
dashed lines, while the solid lines show the effects of the
neutrino postprocessing for a particular choice of fluence [12]. 
The nuclear physics input into these calculations is precisely
that previously described: GT and first-forbidden cross sections,
with the responses centered at excitation energies consistent
with those found in ordinary, stable nuclei, taking into
account the observed dependence on $|N-Z|$. \\

\begin{figure}[htb]
\psfig{bbllx=-4.0cm,bblly=4.5cm,bburx=18cm,bbury=23.0cm,figure=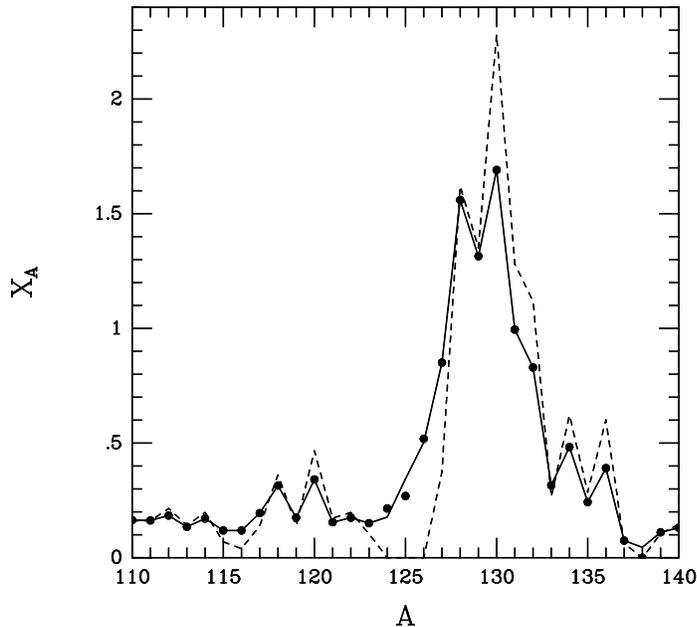,height=3.5in}
\caption{Comparison of the r-process distribution that would 
result from the freezeout abundances near the A $\sim$ 130 
mass peak (dashed line) to that where the effects of neutrino
postprocessing have been include (solid line).  The fluence 
has been fixed by assuming that the A = 124-126 abundances
are entirely due to the $\nu$-process.}
\end{figure}
  
One important aspect of the figures is that the mass shift is
significant.  This has to do with the fact that a 20 MeV 
excitation of a neutron-rich nucleus allows multiple neutrons
( $\sim$ 5) to be emitted.  
(Remember we found that the binding energy of the last neutron
in an r-process neutron-rich nuclei was about 2-3 MeV under
typical r-process conditions.)  The second thing to notice is that
the relative contribution of the neutrino process is particularly
important in the ``valleys" beneath the mass peaks: the reason
is that the parents on the mass peak are abundant, and the
valley daughters rare.  In fact, it follows from this that the neutrino
process effects can be dominant for precisely seven
isotopes (Te, Re, etc.) lying in these valleys.  Furthermore
if an appropriate neutrino fluence is picked, these isotope
abundances are produced perfectly (given the abundance errors).
The fluences are
\[     \mathrm{N} = 82~ \mathrm{peak}~~~~~0.031 \cdot 10^{51} \mathrm{ergs/(100km)^2/flavor} \]
\[     \mathrm{N} = 126~ \mathrm{peak}~~~~0.015 \cdot 10^{51} \mathrm{ergs/(100km)^2/flavor} \]
values in fine agreement with those that would be found
in a hot bubble r-process.  So this is circumstantial but 
significant evidence that the material near the mass cut of 
a Type II supernova is the site of the r-process: there is a
neutrino fingerprint. \\

\begin{figure}[htb]
\psfig{bbllx=-4.0cm,bblly=4.5cm,bburx=18cm,bbury=23.0cm,figure=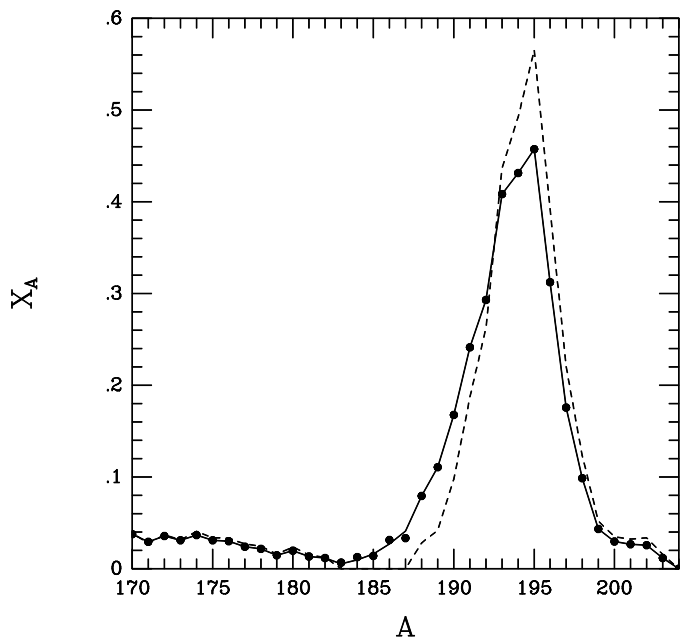,height=3.5in}
\caption{As in Fig. 5, but for the A $\sim$ 195 mass peak.
The A = 183-187 abundances are entirely attributed to the 
$\nu$-process.}
\end{figure}
  
In conclusion, I hope this whirlwind tour through the nuclear
aspects of neutrino interactions in detectors and in stars 
has illustrated how nuclear, neutrino, and stellar physics 
is interconnected.   With the HST and other great observatories
providing so much new information on the nuclear microphysics
governing the universe, it is likely that astrophysics will
continue to provide important challenges to nuclear
physicists.  Likewise, examples like the ``neutrino fingerprint"
on the r-process illustrate how an understanding of nuclear
physics can help astrophysicists settle issues like the site 
of the r-process. \\

The work was supported in part by the US Department of Energy.
  
\section{References}
  
\vspace{1pc}

\re
1.  E. G. Adelberger and W. C. Haxton, 1987, {\it Phys. Rev. C} {\bf 36}, 879.
\re
2.  Austin, S. M., Anantaraman, N., and Love, W. G., 1994,
{\it Phys. Rev. Lett.} {\bf 73}, 30; Watson, J. W. et al., 1985,
{\it Phys. Rev. Lett.} {\bf 55}, 1369.
\re
3.  Bahcall, J. N., 1989, {\it Neutrino Astrophysics} (Cambridge
University Press, Cambridge).
\re 
4.  Bahcall, J. N. and Barnes, C. A., 1964, {\it Phys. Lett}
{\bf 12}, 48.
\re
5.  Bethe, H. and Wilson, J. R., 1985, {\it Ap. J.} {\bf 295}, 14.
\re
6.  Clayton, D. D., 1968, {\it Principles of Stellar Evolution
and Nucleosynthesis} (McGraw-Hill, New York).
\re
7.  Cleveland, B. T. et al., 1998, {\it Ap. J.} {\bf 496}, 505. 
\re
8.  Cooperstein, J., Bethe, H. A., and Brown, G. E., 1984,
{\it Nucl. Phys.} {\bf 429}, 527.
\re
9.  Fuller, G. M. and Meyer, B. S., 1995, {\it Ap. J.} {\bf 453}, 792.
\re
10. Garcia, A. et al., 1991, {\it Phys. Rev. Lett.} {\bf 67}, 3658
and 1990, {\it Phys. Rev. C} {\bf 42}, 765; Trindler, W. et al.,
1995, {\it Phys. Lett. B} {\bf 349}, 267. 
\re
11. Goldhaber, M. and Teller, E., 1948, {\it Phys. Rev.} {\bf 74}, 1046;
Donnelly, T. W., Dubach, J., and Haxton, W. C., 1975,
{\it Nucl. Phys. A} {\bf 251}, 353.
\re
12. Haxton, W. C., Langanke, K., Qian, Y.-Z., and Vogel, P., 1997,
{\it Phys. Rev. Lett.} {\bf 78}, 2694 and {\it Phys. Rev. C}
{\bf 55}, 1532.
\re
13. Mezzacappa, A. et al., 1998, {\it Ap. J.} {\bf 495}, 911;
Janka, H.-Th. and Muller, E., 1996, {\it Astron. Astrophys.} {\bf 306}, 167;
Burrows, A., Hayes, S., and Fryxell, B. A., 1995, {\it Ap. J.}
{\bf 450}, 830.
\re
14. Nakayama, K., Galeao, A. P., and Krmpotic, K., 1982,
{\it Phys. Lett. B} {\bf 114}, 217; D. J. Horen et al., 1980,
{\it Phys. Lett. B} {\bf 95}, 27. 
\re
15. Pfeiffer, B. et al., 1998, astro-ph/9812414.
\re
16. Poskanzer, A. M., McPherson, R., Esterlund, R. A., and
Reeder, P. L., 1966, {\it Phys. Rev} {\bf 152}, 995;
Sextro, R. G., Gough, R. A., and Cerny, J, 1974, 
{\it Nucl. Phys. A} {\bf 234}, 130.
\re
17. Rapaport, J. et al., 1981, {\it Phys. Rev. Lett.} {\bf 47}, 1518.
\re
18. Timmes, F. X., Woosley, S. E., and Weaver, T. A., 1995,
{\it Ap. J. Suppl.} {\bf 98}, 617.
\re
19. Walecka, J. D., 1975, in {\it Muon Physics}, ed. V. W. Hughes
and C. S. Wu (Academic Press, New York), vol. 2, p. 113.
\re
20. Woosley, S. E. et al., 1994, {\it Ap. J.} {\bf 433}, 229;
Woosley, S. E. and Hoffman, R. D., 1992, {\it Ap. J.} {\bf 395}, 202.
\re
21. Woosley, S. E. and Haxton, W. C., 1988, {\it Nature} {\bf 334}, 45;
Woosley, S. E., Hartmann, D. H., Hoffman, R. D., and Haxton, W. C.,
1990, {\it Ap. J.} {\bf 356}, 272.
  
\end{document}